\def\year{2017}\relax
\documentclass[letterpaper]{article}
\usepackage{amsmath}
\usepackage{graphicx}
\usepackage{balance}
\usepackage{color}
\usepackage{url}
\usepackage{subfig}
\usepackage{aaai17}
\usepackage{times}
\usepackage{helvet}
\usepackage{courier}
\usepackage{tabularx}
\usepackage{dblfloatfix}

\newcommand \qa{Q\&A}
\newcommand \ckps{CKPS}
\newcommand \ckc{collaborative knowledge creation}

\begin{document}
%
\title{Dynamics of Content Quality in Collaborative Knowledge Production}
\author{Emilio Ferrara,$^1$ Nazanin Alipourfard,$^1$ Keith Burghardt,$^{1,2}$ Chiranth Gopal,$^{1,3}$ Kristina Lerman$^1$\\
$^1$ University of Southern California, Information Sciences Institute, Marina del Rey, CA (USA) \\
$^2$ Dept of Computer Science, University of California at Davis, Davis, CA (USA)\\
$^3$ PES Institute of Technology, Karnataka (India)
}
%
%
%
\maketitle
\begin{abstract}
We explore the dynamics of user performance in collaborative knowledge production by studying the quality of answers to questions posted  on Stack Exchange. We propose four indicators of answer quality: answer length, the number of code lines and hyperlinks to external web content it contains, and whether it is accepted by the asker as the most helpful answer to the question. Analyzing millions of answers posted over the period from 2008 to 2014, we uncover regular short-term and long-term changes in quality. In the short-term,
quality deteriorates over the course of a single session, with each successive answer becoming shorter, with fewer code lines and links, and less likely to be accepted. In contrast, performance improves over the long-term, with more experienced users producing higher quality answers. These trends are not a consequence of data heterogeneity, but rather have a behavioral origin. Our findings highlight the complex interplay between short-term deterioration in performance, potentially due to mental fatigue or attention depletion, and long-term performance improvement due to learning and skill acquisition, and its impact on the quality of user-generated content.
\end{abstract}

%
%
%
%
%

\noindent  Online collaboration has transformed  how people create knowledge, from general-purpose 
encyclopedias, such as Wikipedia~\cite{kittur2008harnessing},  to 
curation of books and other cultural products~\cite{aiello2010link,mcauley2013amateurs,danescu2013no}. 
Question answering (\qa) sites, such as \textit{Quora}, \textit{Yahoo! Answers}, and \textit{Stack Exchange}, represent an important category of collaborative knowledge production systems (\ckps). On these sites, millions of people ask questions on a multitude of topics, as others answer them asynchronously. Most {\qa} sites integrate a number of features for enhancing \ckc: in addition to asking and answering questions, people can curate both questions and answers by tagging them with descriptive keywords, and identifying the best answers. Curated answers offer a lasting value to the community~\cite{anderson2012discovering}, as they enable future users to quickly find the most helpful answers to their questions. By reducing the time it takes people to find solutions to problems, these sites serve to enhance productivity and accelerate innovation.
One immediate question arises: How good is the knowledge that is collectively produced by an online community? Several researchers tried to address this question by examining the quality of individual contributions and collective outcomes.
Wikipedia, for example, has been subject of extensive study: as one of the prominent \ckps, content quality and methods to assess it have been a central focus of investigation starting a decade ago~\cite{dondio2007computational,kittur2008harnessing,wohner2009assessing,liu2011does}.
%
Early studies focusing on content quality left unchecked the role of users producing that content. Leskovec and collaborators filled this gap by studying the evolution of user behavior in \ckps~\cite{mcauley2013amateurs,danescu2013no}. Their analysis revealed that users of \ckc~platforms change their behavior with experience, and common patterns of evolution emerge over time, which in turn affect perceived content value and objective quality.
New evidence suggests that cognitive dynamics shape human activity on digital platforms: the effect of limited attention on content consumption~\cite{weng2012competition,hodas2012visibility}, and the role of cognitive heuristics in information search and retrieval~\cite{craswell2008experimental,galesic2008eye,Gallottie1500445} are just two examples of such recently discovered phenomena.
The research community just started studying the role of cognitive limits on \ckps.
A study by Singer and collaborators noted a decrease in the quality of comments produced by users
over the course of their activity sessions on Reddit~\cite{singer2016evidence}: sessions of increasing length were associated with shorter, progressively simpler comments, which received declining scores and generated fewer responses from others. This suggests a link between cognitive factors and the dynamics of peer production platforms, specifically the effects of user performance deterioration.
An analysis of voting for best answers on Stack Exchange showed that collective performance is compromised by individual-level cognitive biases and response to cognitive load~\cite{burghardt2016myopia}.

\vspace*{-2mm}
\subsubsection{Contributions of this work.} We explore the
behavioral factors affecting the quality of user-generated content by studying a  data set containing  millions of answers posted on Stack Exchange during the period  2008--2014.
To control for behavioral heterogeneity, we segment
user activity into sessions---sequences of answers written by the same user without  an extended break. This allows us to compare users who expend similar levels of effort, thereby reducing some of the individual variability.

\noindent Our work addresses the following research questions:

\smallskip
\noindent{\textbf{RQ1}:} What are the short-term changes in 
the quality of content users produce over the course of a single session? To this end, we capture the quality of answers posted on Stack Exchange by means of quantitative indicators, such as their length, probability of acceptance,
number of hyperlinks contained in the answer, as well as number of lines of code therein included.

\smallskip
\noindent{\textbf{RQ2}:} What are the long-term changes in 
the quality of content users produce? Does user tenure---a combination of accumulating experience and learning dynamics---affect 
their performance? What role does user tenure play in short-term changes in performance? Do 
novice and veteran users produce different quality content, and does their behavior change the same way over the course of sessions?.

\smallskip By addressing these questions, we will shed light on a new and untamed issue in \ckps, namely the short-term deterioration in user contribution's quality (associated with mental fatigue and attention depletion), as well as describe the effects of learning dynamics and long-term platform adoption. Understanding performance dynamics will pave way for the next generation of intelligent user interfaces that monitor and predict performance, and maybe intervene at the right time so as to maximize human performance. Such performance gains could yield substantial benefits: even small individual improvements would result in long-term benefits of higher quality knowledge systems.

\subsubsection{Data.}
Stack Exchange launched in 2008 as a place for asking programming questions. It has grown vigorously, adding more forums on a variety of technical and non-technical topics.
The premise behind Stack Exchange is simple: any user can ask a question, which others may answer. Users can also {vote} for answers they find helpful, and the asker can {accept} one of the answers as the best answer to the question. Stack Exchange highlights accepted answers and those with most votes, making it easy for others to find them. 
We used anonymized data representing all questions and answers from August 2008 until September 2014 (\textit{https://archive.org/details/stackexchange}). The data includes
9.6M questions, of which approximately half had an accepted answer.
Only the questions that received two or more answers were included in our study.  
This step helped filter out answers that were trivially accepted because they were the only answers users saw.
We also recorded user attributes, including the time of user sign-up.

\subsubsection{Answer Quality.}
To answer the research questions, we need a measure of answer quality. In general, this is a complex and often subjective issue, making it difficult to quantify.
However, we have reasonable expectations for what makes a good answer: better answers tend to be more extensive (i.e., contain more words), they provide examples of solutions to the question (e.g., include code snippets), and support the argument with external references to documentation or other resources (e.g., contain hyperlinks to external Web content), and finally, they are judged as helpful answers (i.e., accepted by the asker).
We use these heuristics to define quantitative indicators to serve as proxies for answer quality:

\noindent\textbf{i) Acceptance Probability:} 
the probability that the asker selects the answer as the most helpful to the asker personally.

\noindent\textbf{ii) Number of Words:} 
the size of the body of the post (i.e., after removing URLs and programming code).

\noindent\textbf{iii) Number of Lines of Code:} 
accounts for snippets of code potentially included in the answer.

\noindent\textbf{iv) Number of Links:} 
accounts for the number of URLs pointing to external resources users include in their answers.

\subsubsection{Sessions.}
Some people are able to devote more effort to answering questions on Stack Exchange than others. To partially account for individual variability, we segment user activity into sessions, periods of continuous activity without a prolonged break, usually characterized by a single intent~\cite{jones2008beyond,huang2009analyzing}.
To construct sessions from the time series of user activity, we examined the time interval between consecutive answers posted by the same user.
The distribution had a peak at short time scales (10-20 minutes) and very long ones ($>$1000 minutes, i.e., one day) suggesting activity is affected by short-term changes and daily routines. A  wide valley appears between these two, which suggests that any choice of the threshold in this range should yield mostly equivalent results: we select 100 minutes as the threshold time interval that defines activity sessions. A break longer than 100 minutes constitutes the end of a session.
Variations to this parameter leave the results below essentially unchanged.
Using 100 minute threshold, we segment user activity into sessions and measure the length of each session, defined as the number of answers user produced within that session.
Most activity sessions are short: 73.4\% contain only one answer, 
and 97\% of all sessions contain five or fewer answers. 

%




\begin{figure*}[t]\centering
\includegraphics[width=.52\columnwidth,height=.52\columnwidth]{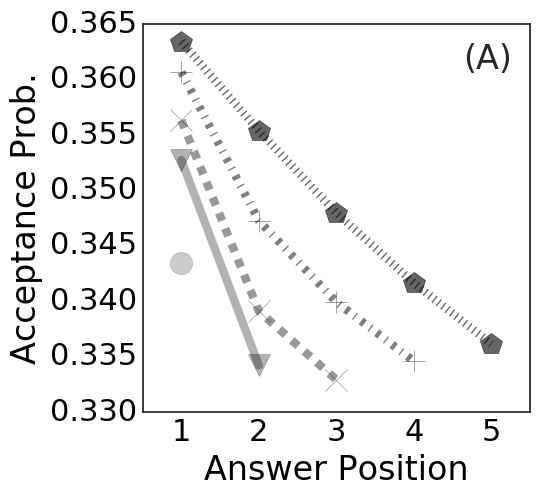}
\includegraphics[width=.52\columnwidth,height=.52\columnwidth]{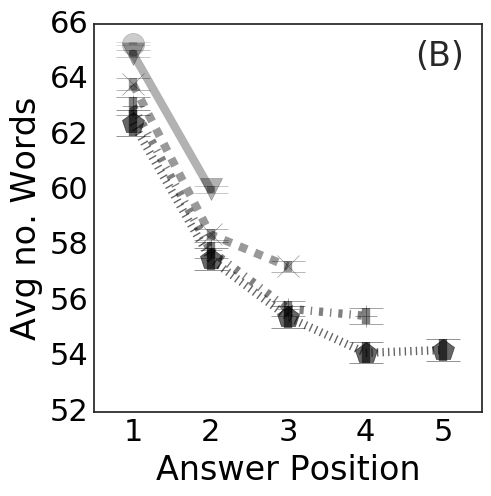}
\includegraphics[width=.52\columnwidth,height=.52\columnwidth]{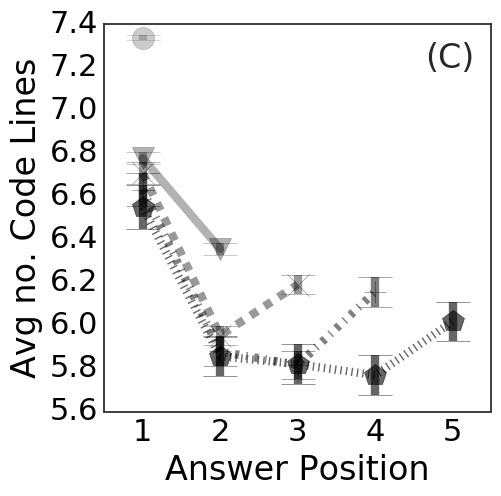}
\includegraphics[width=.52\columnwidth,height=.52\columnwidth]{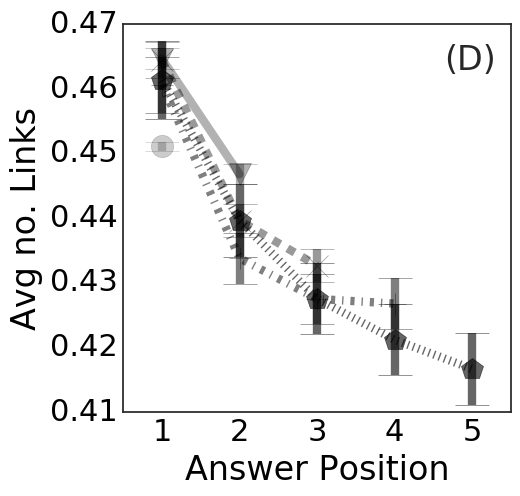}\vspace{-.4cm}
\caption{\textit{Answer quality deteriorates over  the course of sessions of different length}. B-D also report 95\% confidence intervals.
}
\label{fig2}\vspace{-.5cm}
\end{figure*}

\subsection{RQ1: Short-term Performance Dynamics}
Figure~\ref{fig2} shows that answer quality deteriorates over the course of a session,
with panels A--D reporting values of the four indicators of answer quality as a function of answer's position within the session of a given length.  The five lines correspond to sessions of length one to five. For example, in Fig.~\ref{fig2}A,
the gray dash-dotted line shows average acceptance probability of answers in sessions where four answers were written by a user: the first answer of such sessions is accepted by the asker about 36\% of the time, while the last answer is accepted less than 33.5\% of the time.
Similar declines in quality are evident across sessions of all lengths, and across all quality indicators. All declines are statistically significant (95\% confidence intervals are often obscured by the marker).
Figure~\ref{fig2}A suggests that in sessions where multiple answers are written, acceptance probability decreases about 10\% between the first and the last answer of the session, highlighting a somewhat large short-term decline in acceptance probability.
Figure~\ref{fig2}B shows the length of the answer as a function of its position within a session. Here the decline is even more pronounced, suggesting that consecutive answers become ever shorter, with a difference of about 20\% between the length of the first and last answer. A decline on the order of 10\% is visible also in the number of code lines  (Fig.~\ref{fig2}C) and hyperlinks (Fig.~\ref{fig2}D) provided in an answer.
Another  effect in Fig.~\ref{fig2} is the stacking of session performance. For example, answers posted during longer sessions are more likely to be accepted than answers posted during shorter sessions (Fig.~\ref{fig2}A). A similar stacking was observed in \cite{singer2016evidence}. Such behavior could be explained by the fact that high reputation users answer questions first~\cite{anderson2012discovering}. Presumably, such users are interested in improving their reputation and look for questions to answer, and as a result, write more answers during a session. Since these users are more experienced, as we show later, they produce higher quality answers, possibly explaining the stacking of probability of acceptance curves.

\paragraph{Randomized sessions.} Quality deterioration may be an artifact of data heterogeneity. To test this hypothesis, we designed a null model that disrupts sessions by randomizing the time interval between the answers~\cite{singer2016evidence}. In the \emph{randomized session data}, we shuffled the time intervals  between consecutive answers written by a given user, but preserved all the other features, including the temporal order of answers. Then, we simply  segmented user activity into sessions based on  randomized times. This randomization removed any short-term performance decline in the reshuffled data, corroborating the hypothesis that 
answers written later in a session are of lower quality.


\begin{figure*}[t]\centering
\includegraphics[width=.52\columnwidth,height=.52\columnwidth]{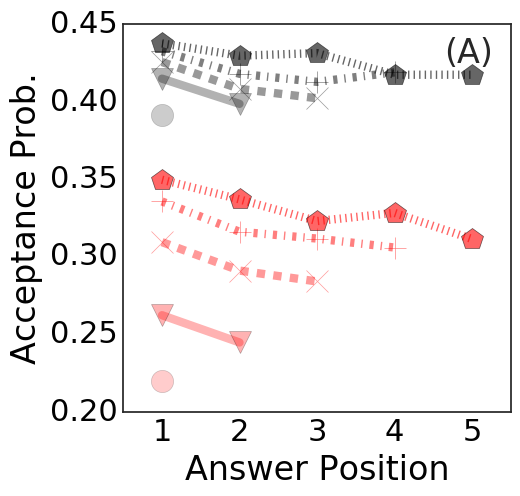}
\includegraphics[width=.52\columnwidth,height=.52\columnwidth]{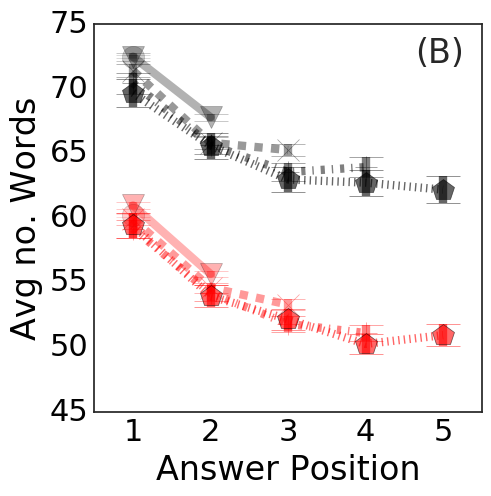}
\includegraphics[width=.52\columnwidth,height=.52\columnwidth]{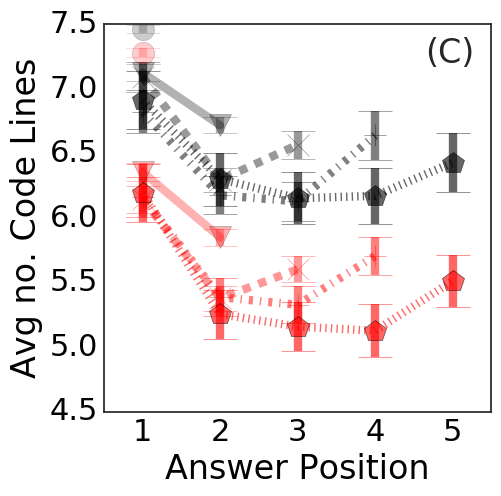}
\includegraphics[width=.52\columnwidth,height=.52\columnwidth]{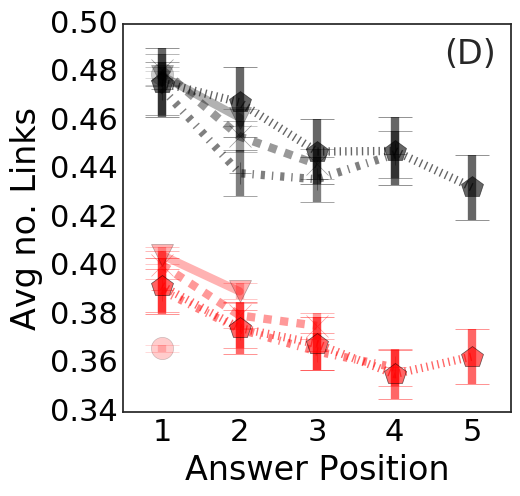}\vspace{-.4cm}
\caption{\textit{Short-term decline vs. tenure}: Veterans (black) systematically exhibit higher average performance than novices (red). 
}
\label{fig6}\vspace{-.3cm}
\end{figure*}

\begin{figure*}[t]\centering
\includegraphics[width=.52\columnwidth,height=.52\columnwidth]{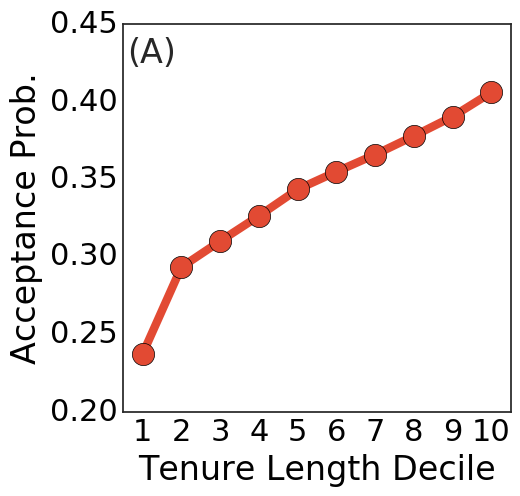}
\includegraphics[width=.52\columnwidth,height=.52\columnwidth]{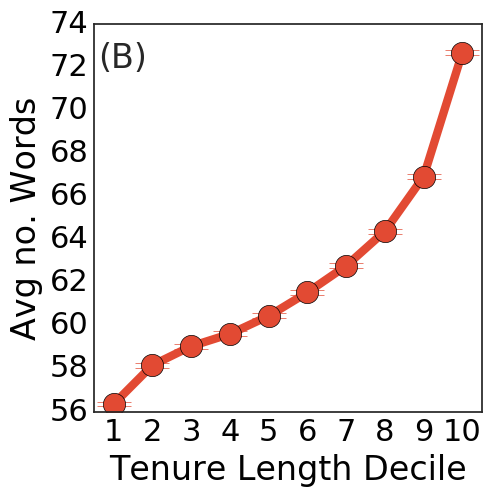}
\includegraphics[width=.52\columnwidth,height=.505\columnwidth]{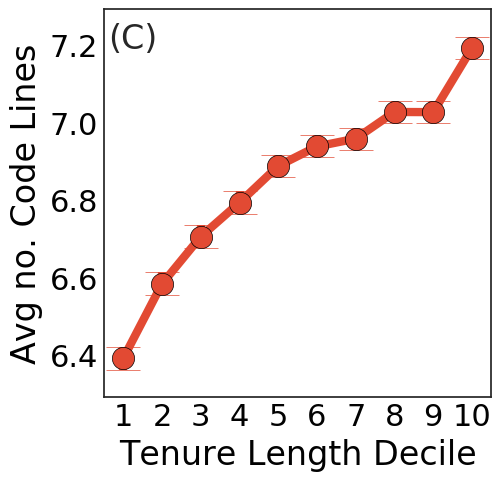}
\includegraphics[width=.52\columnwidth,height=.52\columnwidth]{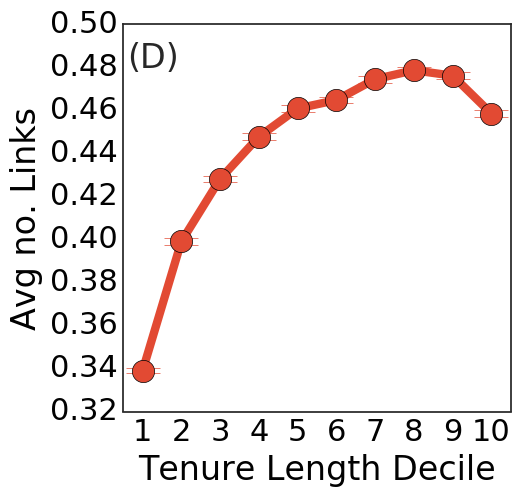}\vspace{-.4cm}
\caption{\textit{Long-term performance increase vs. tenure length}. Plots B-D report 95\% confidence intervals. 
}
\label{fig7}\vspace{-.5cm}
\end{figure*}

\subsection{RQ2: Long-term Performance Dynamics}

\subsubsection{Novices vs Veterans.}  We hypothesize that the combination of accumulating experience and learning can affect the quality of content produced by users with longer tenure.
To compute the length of user $u_i$'s tenure at the time $t_{ij}$ the user composed the answer $a_{ij}$, we take the difference, in months, from the time $u_i$ posted his or her very first answer: $\tau_j(u_i)=t_{ij}-t_{i1}$. This tells us how long the user has been active on Stack Exchange at the time the answer was posted.
For all answers in our data set, we then calculate user tenure length $\tau_j(u_i)$ at the time each answer was created. 
20\% of the answers were written 
\emph{novice} who have been active on Stack Exchange less than six months; 60\% of the answers were written by users who have been active for less than 27 months; and 80\% of the answers were written by users who have been active for less than 42 months. The remaining 20\% of the answers written by most experienced, or \emph{veteran}, users. 
Figure~\ref{fig6} reports the quality of answers written by novices and veterans. 
We observe within-session performance declines  for both novices (shades of red lines) and veterans (shades of grey lines) similar to Fig.~\ref{fig2}.
However, all four answer quality indicators are significantly higher for veterans than novices, supporting the hypothesis  that tenure length affects performance. For example, approximately 39\% of answers written by veteran users during sessions of length one are accepted, compared to just 22\% of answers written by novices during similar sessions. This is a significant difference of more than 40\%. 
Tenure length also seems to affect other performance metrics: the average number of words for novices starts at about 60 words per answer for the first answer of each session, whereas that of veterans is about 72 words, a difference of 20\%. This difference between novices and veterans in the order of 20\% is present also for the average number of lines of code in an answer, and the average number of provided links.
In fact, the last answer written by a veteran at the end of a long question-answering sessions is typically better than the first answer written by a novice user, who has not yet experienced effects of performance deterioration.
Remarkably, short-term performance deterioration of veterans is very similar to that of novice users, suggesting that depletion is governed by mechanisms that are not affected by user experience, learning, or user reputation, but are likely linked to intrinsic cognitive limits, such as limited attention and the effect of fatigue.

\vspace*{-2mm}
\subsubsection{Performance vs. Tenure.}
We divided all answers into deciles based on the tenure of their authors at the time the answers were posted.
Accordingly, answers in the first decile represent the 10\% of the answers
written by ``youngest'', least experienced users (who joined the platform most recently), while the tenth decile represents the 10\% of the answers written by ``oldest'' veterans (who have been on Stack Exchange longest).
Figure~\ref{fig7} shows the four quality indicators as a function of tenure length deciles. There exists a positive trend with tenure length in all plots, suggesting that longer tenure is associated with better performance and higher quality answers. For example, only one in four answers written by users in the first tenure deciles is accepted versus 40\% of the answers written by users in the top tenure decile. This trend is also evident for the average number of words: more experienced users produce longer answers (over 70 words on average, as opposed to less than 60 for users with less experience). The same applies to the average number of lines of code associated with answers, and for average number of links, with a difference on the order of 20\% between first and tenth deciles.
This  corroborates the hypothesis that  experience improves performance.

\subsection{Conclusions}
We have explored dynamics of quality of user-generated content in a collaborative knowledge production system by analyzing millions of answers posted on Stack Exchange. As a proxy of quality, we used four quantitative indicators and studied how these change over the course of user activity on the platform.
In the short term, i.e., over the course of a single session, content quality declines substantially. As this performance deterioration is similar to that observed recently on other platforms, such as Reddit~\cite{singer2016evidence} and Twitter~\cite{kooti2016twitter}, we suspect that it has a cognitive origin---e.g., due to mental fatigue, loss of attention, or boredom. Further work is necessary to investigate this connection.
Over the long term, however, overall users' performance improves: this is potentially due to a combination of factors  such as learning, emergence of expertise, and skill acquisition, that affect veteran users; on the flip side of the coin, poorly performing users tend to drop out.

Our work raises the possibility of new assessment tools that could improve human performance and the quality of knowledge production systems in general. Such cognitive assessment tools could monitor individual behavior and intervene at the right time, for example,
by suggesting a break, so as to optimize online performance and user experience. 
Our long-term research plan includes providing new strategies for the design of personalized incentive mechanisms to enhance user experience in online collaborative platforms.

\vspace*{-2mm}
\paragraph{Acknowledgements.} This work was supported in part by DARPA (\#D16AP00115), NSF (\#SMA-1360058), ARO (\#W911NF-15-1-0142), and the USC Viterbi India summer research program.
No official endorsement nor reflection of the US Government's position/policy should be inferred. Approved for public release: unlimited distribution.


\end{document}